\documentclass[epjCONF]{svjour}
\usepackage{graphics,graphicx}
\usepackage[varg]{txfonts} % Times fonts
\usepackage[latin1]{inputenc}
\newcommand{\HST}{{\sl HST}}
\newcommand{\Msun}{\mbox{$M_{\odot}$}}
\newcommand{\Lsun}{\mbox{$L_{\odot}$}}
\newcommand{\Mjup}{\mbox{$M_{\rm Jup}$}}
\newcommand{\Mtot}{\mbox{$M_{\rm tot}$}}
\newcommand{\Lbol}{\mbox{$L_{\rm bol}$}}
\newcommand{\Teff}{\mbox{$T_{\rm eff}$}}
\newcommand{\Mstar}{\mbox{$M_{\star}$}}
\session-title{New Technologies for Probing the Diversity of Brown
  Dwarfs and Exoplanets}
\begin{document}
\title{Testing Substellar Models with Dynamical Mass Measurements}
\author{Trent J.\
  Dupuy,\inst{1}\fnmsep\thanks{\email{tdupuy@ifa.hawaii.edu}} Michael
  C.\ Liu,\inst{1} \and Michael J.\ Ireland\inst{2}}
\institute{Institute for Astronomy, University of Hawai`i, 2680
  Woodlawn Drive, Honolulu, HI 96822 \and School of Physics,
  University of Sydney, NSW 2006, Australia}
\abstract{ We have been using Keck laser guide star adaptive optics to
  monitor the orbits of ultracool binaries, providing dynamical masses
  at lower luminosities and temperatures than previously available and
  enabling strong tests of theoretical models.  We have identified
  three specific problems with theory: (1)~We find that model
  color--magnitude diagrams cannot be reliably used to infer masses as
  they do not accurately reproduce the colors of ultracool dwarfs of
  known mass.  (2)~Effective temperatures inferred from evolutionary
  model radii are typically inconsistent with temperatures derived
  from fitting atmospheric models to observed spectra by 100--300~K.
  (3)~For the only known pair of field brown dwarfs with a precise
  mass (3\%) \emph{and} age determination ($\approx$25\%), the
  measured luminosities are $\sim$2--3$\times$ higher than predicted
  by model cooling rates (i.e., masses inferred from \Lbol\ and age
  are 20--30\% larger than measured).  To make progress in
  understanding the observed discrepancies, more mass measurements
  spanning a wide range of luminosity, temperature, and age are
  needed, along with more accurate age determinations (e.g., via
  asteroseismology) for primary stars with brown dwarf binary
  companions.  Also, resolved optical and infrared spectroscopy are
  needed to measure lithium depletion and to characterize the
  atmospheres of binary components in order to better assess model
  deficiencies.}
\maketitle

\section{Introduction}
\label{intro}

Detailed theoretical models of stars, developed and observationally
tested over the last century, now underlie most of modern astronomy.
However, the lowest mass stars ($\Mstar \lesssim 0.1 \Msun$) are
sufficiently cool ($\Teff \lesssim 3000$~K) that the standard,
well-tested stellar models are not appropriate (e.g., due to dust
formation in the photosphere). Objects below the hydrogen-fusing mass
limit can cool to even lower temperatures as they have no sustained
source of internal energy generation. At temperatures below
$\sim$2000~K, dust and H$_2$O are the major sources of opacity in the
photosphere, and below $\sim$1400~K methane absorption becomes
important. Over the last decade, new theory has been developed to
describe such low-temperature objects, which encompass brown dwarfs
and gas-giant planets (e.g., \cite{2001ApJ...556..357A},
\cite{2003A&A...402..701B}, \cite{1997ApJ...491..856B},
\cite{2005astro.ph..9066B}, \cite{2000ApJ...542..464C},
\cite{2008ApJ...678.1419F}). These models now form the basis of our
understanding of all low-mass gaseous objects, from stars at the
bottom of the main sequence to extrasolar giant planets. Thus,
rigorously testing them is vitally important.

Dynamical masses for ``ultracool'' visual binaries (i.e., those with
spectral types later than $\sim$M7) are central to this effort, but
such measurements have previously been impeded by observational
limitations.  Most ultracool binaries were discovered $\sim$5 to
10~years ago, and the shortest estimated orbital periods are long
($\gtrsim$10--30~years). The corresponding physical separations
($\sim$1--3~AU) result in very small angular separations
($\lesssim0.2$ arcsec) that can only be resolved using the
\textsl{Hubble Space Telescope} (\HST) or ground-based adaptive optics
(AO).  Direct distance measurements via trigonometric parallax are
also essential as they are needed to convert the observed angular
scale of the orbit into physical units and equally importantly to
provide a direct measurement of the luminosity.  As a result, previous
to our work only three ultracool binaries had measured dynamical
masses (\cite{2004A&A...423..341B}, \cite{2001A&A...367..183L},
\cite{2004astro.ph..7334O}), and these were all relatively warm
($\gtrsim2100$~K).

\section{Dynamical Masses of Ultracool Binaries}
\label{sec:masses}

\begin{figure}
%\centering

\hskip 0.15in
\centerline{
\resizebox{0.75\columnwidth}{!}{\includegraphics[angle=90]{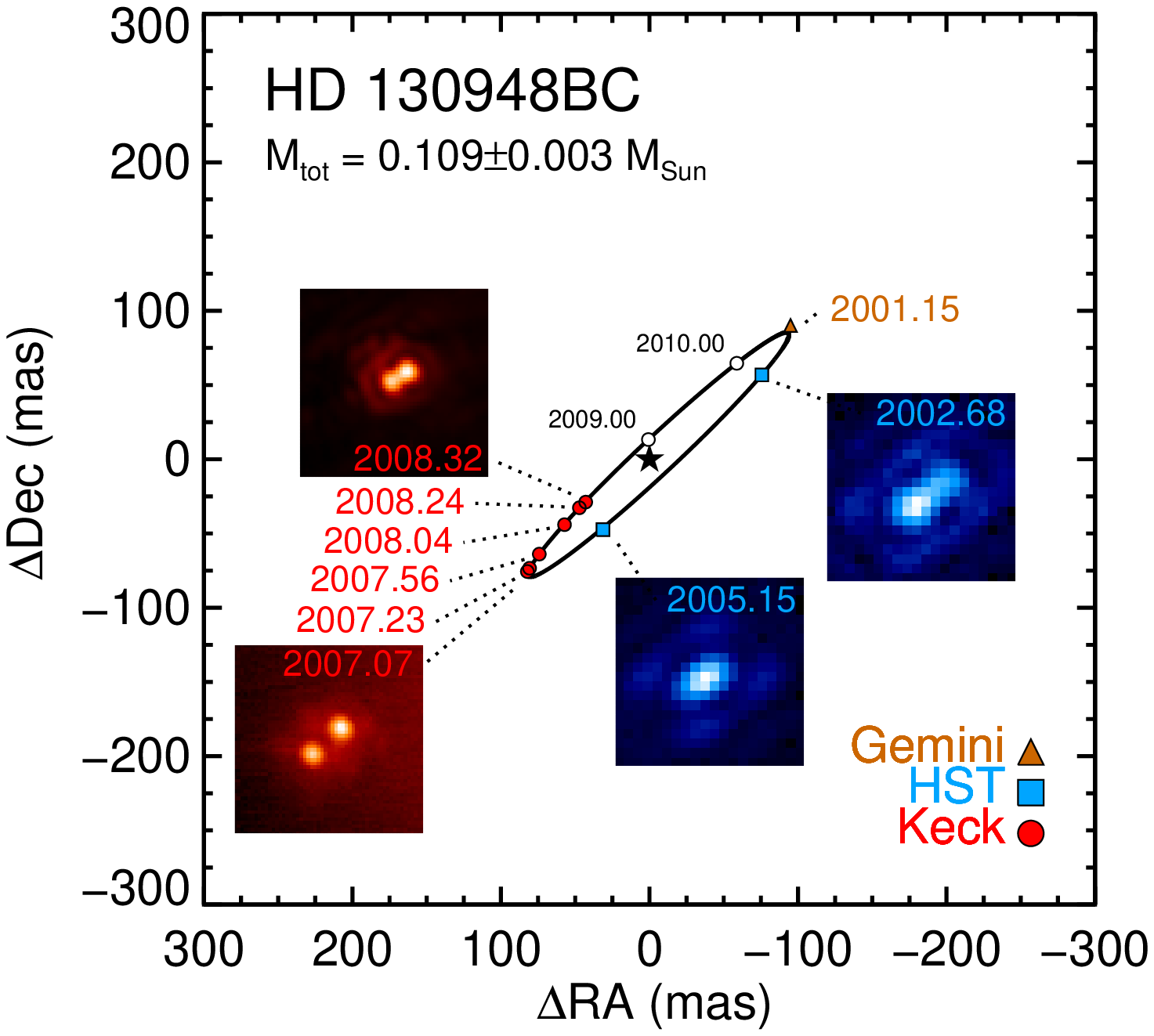}}
\hskip -0.8in
\resizebox{0.75\columnwidth}{!}{\includegraphics[angle=90]{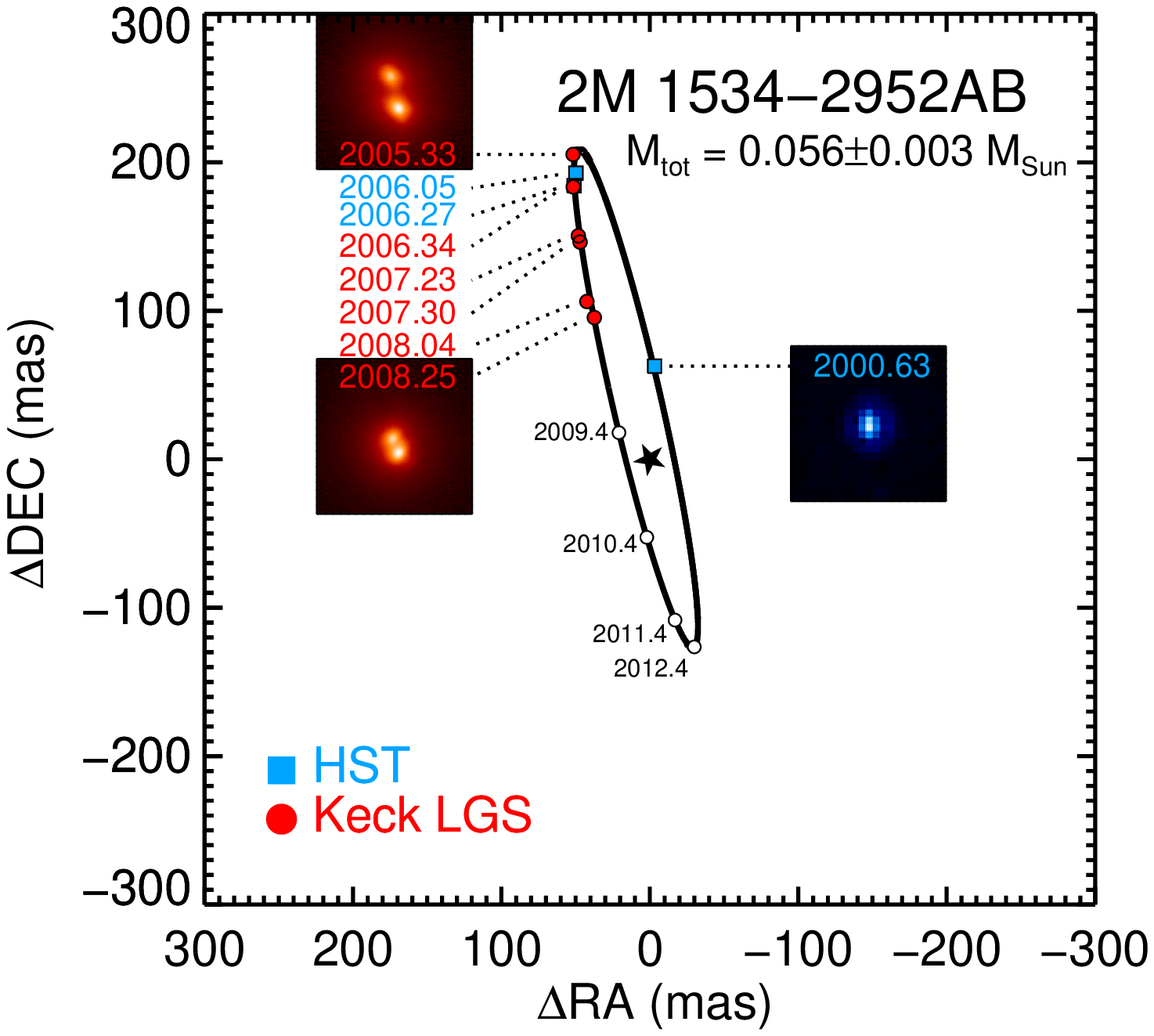}}
}
\centerline{
\resizebox{0.5\columnwidth}{!}{\includegraphics{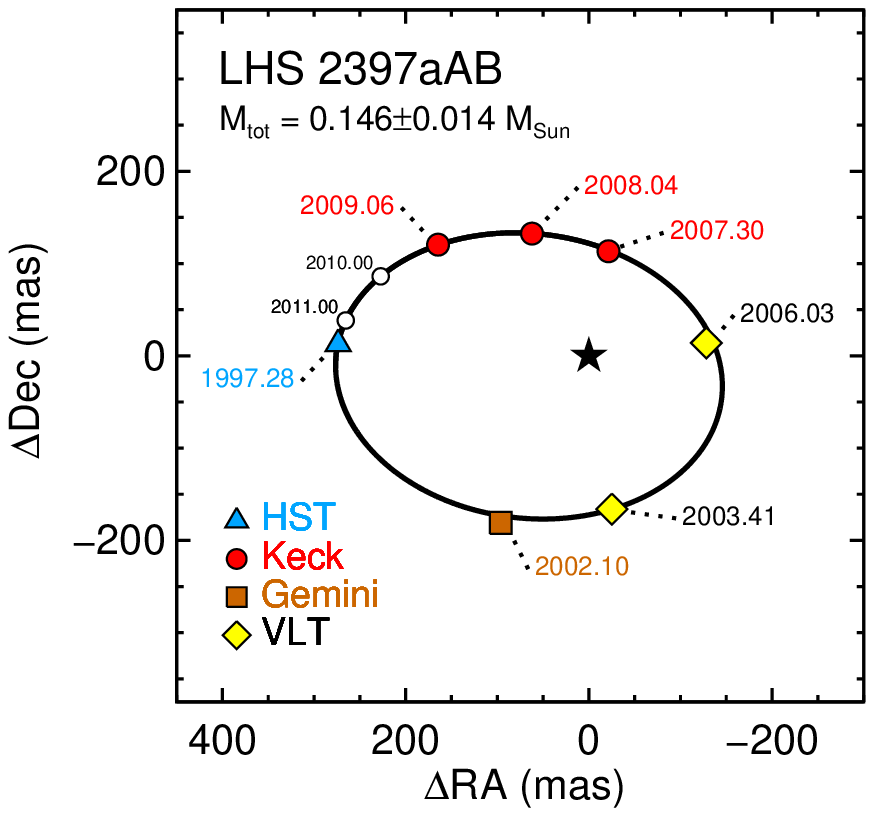}}
\resizebox{0.5\columnwidth}{!}{\includegraphics{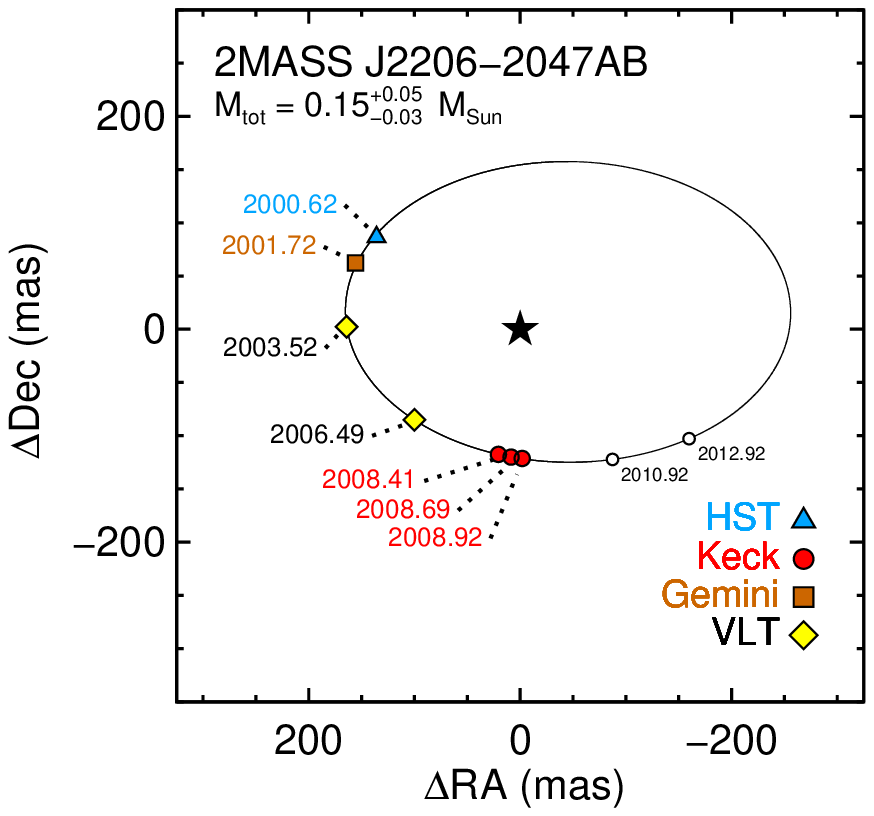}}
}

\caption{ Our Keck LGS AO data combined with discovery and archival
  data from 5 to 10 years ago enables precise orbit determinations for
  ultracool binaries. {\em Top left:} HD~130948B and C are companions
  to a young solar analog (G2V, $0.8\pm0.2$~Gyr), making them the
  first field brown dwarfs (L4+L4) with a well-determined age
  \emph{and} masses ($\Mtot = 0.109\pm0.003$~\Msun;
  \cite{2009ApJ...692..729D}).  {\em Top right:} 2MASS~J1534$-$2952AB
  is the first T~dwarf binary with a dynamical mass (T5+T5.5, $\Mtot =
  0.056\pm0.003$~\Msun), revealing inconsistencies between the
  atmospheric model-derived temperatures, evolutionary model H--R
  diagram, and measured mass \cite{2008ApJ...689..436L}.  {\em Bottom
    left:} LHS~2397aAB (M8+L7, $\Mtot = 0.146\pm0.014$~\Msun) is the
  first dynamical mass benchmark at the L/T transition, showing
  consistency between temperatures estimated from atmospheric and
  evolutionary models and supporting the idea that the temperature of
  the L/T transition is surface gravity dependent \cite{me-2397a}.
  {\em Bottom right:} 2MASS~J2206$-$2047AB (M8+M8) is a pair of stars
  at the bottom of the main sequence that have $J$-band colors
  0.2--0.3~mag redder than predicted by evolutionary model tracks for
  objects of their measured masses \cite{me-2206}.  This implies that
  masses and/or ages inferred from model color--magnitude diagrams
  will be in error for such objects.}
\label{fig:orbit}
\end{figure}

We have been using Keck laser guide star (LGS) AO direct imaging and
aperture masking to monitor the orbits of ultracool binaries, enabling
dynamical mass measurements for the lowest mass (30--75~\Mjup), lowest
temperature (1000--2800~K), lowest luminosity ($10^{-4}$ to
$10^{-5}$~\Lsun) objects known.  This has allowed models to be tested
in the unexplored area of parameter space shared by brown dwarfs and
extrasolar giant planets.  Keck LGS AO delivers diffraction-limited
imaging in the infrared for targets over most of the sky ($\sim$0.05''
FWHM; $\sim$4$\times$ sharper than \HST), so it is ideally suited for
resolving short-period ultracool binaries. By performing a detailed
analysis of these high-resolution images and accounting for small
astrometric shifts due to differential atmospheric refraction, we
routinely achieve sub-milliarcsecond astrometric accuracy and
$\sim$200~$\mu$as for our best data. Such high quality astrometry has
allowed us to precisely measure binary orbits, with the error in the
resulting masses (typically 3--10\%) dominated by the uncertainty in
the distance (Figure~\ref{fig:orbit}).

We have also undertaken a substantial amount of supplementary analysis
to enable these mass measurements:

\noindent
(1)~We have re-analyzed archival \HST\ images from 5 to 10 years ago,
improving astrometric errors by a factor of 2 to 8 compared to
published values, and this has proved critical for accurate orbit
determination (e.g., \cite{me-2397a}, \cite{2008ApJ...689..436L}).

\begin{figure}

\centerline{
\resizebox{0.5\columnwidth}{!}{\includegraphics{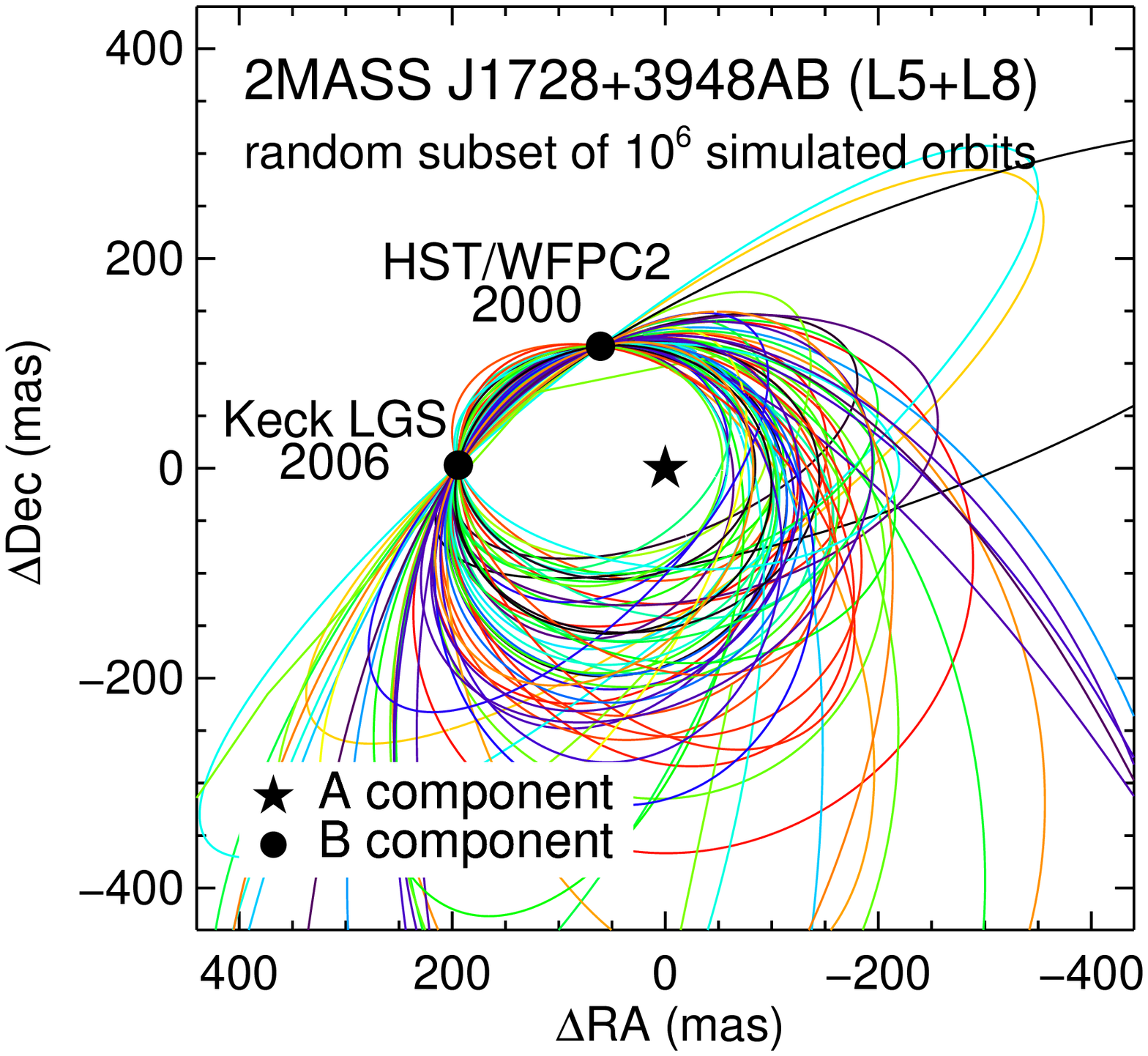}}
\resizebox{0.5\columnwidth}{!}{\includegraphics{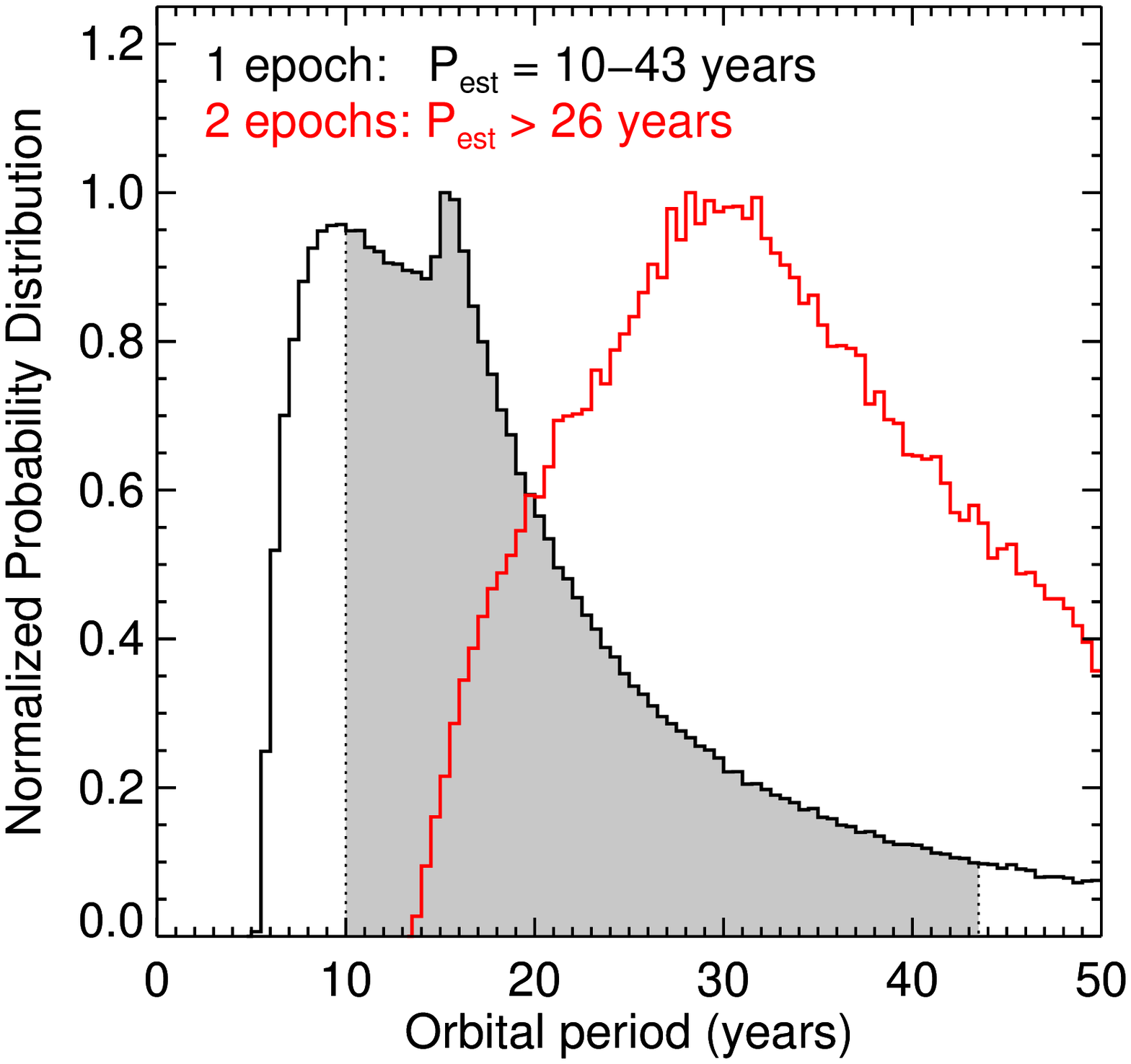}}
}

\caption{We have developed a novel Monte Carlo technique to determine
  the orbital period probability distribution from orbital motion
  observed between only two epochs. {\em Left:} For the ultracool
  binary 2MASS~J1728$+$3948AB, the relative positions of the A and B
  components are shown at the discovery epoch and $\sim$6 years later
  from our Keck LGS AO program (filled symbols). Using the approach
  described in the text, we randomly drew orbits that pass through the
  two observed positions at the appropriate epochs (multi-colored
  lines). {\em Right:} The orbital period distribution of the randomly
  drawn orbits (red) compared to the estimated period distribution
  using only the discovery epoch (black), following the method of
  Torres \cite{torres99}, with $\pm1\sigma$ confidence limits shaded
  gray. Although the two-epoch distribution appears at face value to
  be broader and thus less precise, it is actually strongly preferred
  as it is free of the somewhat arbitrary assumptions required by the
  single-epoch estimate (i.e., a uniform eccentricity distribution
  between $0 < e < 1$ and a total mass of 0.135~\Msun).  In the case
  of 2MASS~J1728$+$3948AB, we found that the orbital period is on the
  longer side ($P > 26$~years, 68.3\% c.l.) of the original estimate
  (10--43~years, 68.3\% c.l.), reducing its priority in our orbital
  monitoring program.} \label{fig:two-epoch}

\end{figure}

\noindent
(2)~We have developed a novel Monte Carlo technique to determine the
orbital period probability distribution from motion observed between
discovery and our first Keck data obtained $\gtrsim5$--$10$ years
later (Figure~\ref{fig:two-epoch}).
This has enabled us to accurately gauge target priorities and thus
measure dynamical masses faster with a limited amount of telescope
time.
The orbital period, and thus the monitoring priority, of a visual
binary is very uncertain from a single observation.  In order to
estimate the period probability distribution from a single
observation, one must assume both a total mass and eccentricity
probability distribution, and even with these assumptions the
$\pm1\sigma$ confidence limits span a factor of $\sim$4 in orbital
period \cite{torres99}.  Our method utilizes the two positions and two
times of two observations taken several years apart to eliminate 4 of
the 7 orbital parameters (the two ``geometrical'' parameters semimajor
axis and eccentricity; and the two ``time'' parameters period and time
of periastron passage).  This leaves only 3 parameters (inclination,
argument of periastron, and position angle of the ascending node),
which are just viewing angles that we conservatively assume to be
distributed randomly according to appropriate distributions.  We use a
Monte Carlo approach that results in an ensemble of orbits that pass
through the two observed positions at the observed epochs.  The
distribution of periods of these orbits is the period probability
distribution.  This method can, but does not necessarily, result in a
narrower range of orbital periods; however, this method always results
in a more accurate estimate of the period because it does not require
an assumption about the eccentricity or total mass.  

\begin{figure}

\resizebox{1.0\columnwidth}{!}{\includegraphics[angle=90]{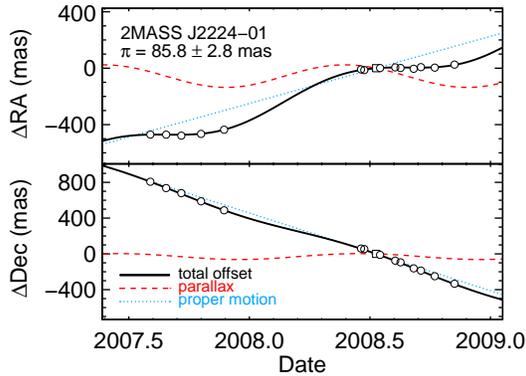}}

\caption{Analysis of data from our infrared parallax program at CFHT
  demonstrates the capability to measure precise distances to visual
  binaries in our Keck LGS AO sample, which is critical as the derived
  dynamical mass depends strongly on the distance ($\Mtot \propto
  d^3$).  Our parallaxes of ``control'' objects that have published
  parallaxes, such as the object shown here, agree well with published
  results.  Our astrometric measurements are shown as open circles
  (error bars are smaller than the plotting symbols) along with the
  best-fit astrometric solution (solid black) which includes parallax
  (red dashed) and proper motion (blue dotted) components.}
\label{fig:plx}
\end{figure}

\noindent
(3)~Finally, and most importantly, we have been conducting an infrared
parallax program at the Canada France Hawaii Telescope (CFHT;
Figure~\ref{fig:plx}).  Precisely measured distances are critical for
dynamical masses from visual binaries given the strong dependence of
$\Mtot \propto d^3$. Only about 1 in 4 of the shortest period
ultracool binaries have previously published parallaxes, so our
program targeting $\sim$30 binaries enables a greatly expanded sample
of masses.

\section{Testing Substellar Models}
\label{sec:test}

Our Keck program has been yielding dynamical masses with the needed
precision (3--10\%) to perform fundamental tests of theory
(Figure~\ref{fig:orbit}).  To date, we have identified three specific
problems with substellar models:

\noindent
(1)~We find inconsistencies between predicted near-infrared colors and
those observed for field objects of known mass over a wide range of
spectral types (\cite{me-2206}, \cite{2009ApJ...692..729D},
\cite{me-2397a}, \cite{2008ApJ...689..436L}).  For example, the M8+M8
components of 2MASS~J2206$-$ 2047AB and L4+L4 components of HD~130948BC
appear to be $\approx$0.2--0.4~mag redder than model tracks, while the
T5+T5.5 components of 2MASS~J1534$-$2952AB appear to be
$\approx$0.2--0.4~mag bluer.  This is likely due to imperfect modeling
of dust for the late-M and L dwarfs and incomplete methane line lists
for the T dwarfs.  Thus, masses and/or ages inferred from theoretical
color--magnitude diagrams should be treated with caution.

\noindent
(2)~If theoretically predicted radii are correct, we have found that
temperatures derived from atmospheric models are systematically in
error (\cite{me-2206}, \cite{2009ApJ...692..729D},
\cite{2008ApJ...689..436L}).  This points either to inaccurate
theoretical radii or to incomplete modeling of such low-temperature
atmospheres.  The one object for which this is not the case is
LHS~2397aB, currently the only mass benchmark at the L/T transition
\cite{me-2397a}.  This is surprising as existing atmospheric models
should not be appropriate for such transitional objects as their
predictions are only valid in the limiting cases of maximal dust
(Dusty, \cite{2000ApJ...542..464C}) and the total absence of dust
(COND, \cite{2003A&A...402..701B}).  Thus, in the case of LHS~2397aB
it is likely that large systematic errors cancel out to produce
apparent agreement.

\begin{figure}

\resizebox{1.0\columnwidth}{!}{\includegraphics[angle=90]{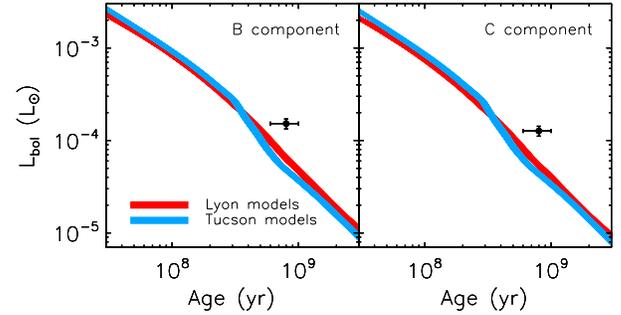}}

\caption{We have tested substellar cooling rates by measuring the
  masses and luminosities of HD~130948B and C, brown dwarfs with a
  well-determined age of 0.8$\pm$0.2~Gyr from the primary star
  HD~130948A \cite{2009ApJ...692..729D}.  Surprisingly, they are
  $\sim$2--3$\times$ more luminous than evolutionary models predict.
  Two independent sets of theoretical luminosity tracks are shown as
  colored isomass lines, where the thickness corresponds to the
  uncertainty in the measured mass.  Such a systematic error would
  imply that model-inferred masses are over estimated by
  $\sim$20--30\% in the usual case where only the luminosity and age
  are known (e.g., for directly imaged extrasolar planets and
  determinations of cluster initial mass functions).  The other
  possibility is that the age of HD~130948A is inaccurately estimated
  despite using the latest gyrochronology calibrations \cite{M&H08}.
  This can be resolved by obtaining a better age estimate (e.g., from
  asteroseismology) and measuring masses and ages for more such
  systems.}
\label{fig:lbol-age}
\end{figure}

\noindent
(3)~For the only system with both a known mass and age, we have found
the measured luminosities of the individual components to be
$\sim2$--$3\times$ higher than predicted (Figure~\ref{fig:lbol-age};
\cite{2009ApJ...692..729D}). This would imply that model-derived
masses are significantly over estimated by $\sim$20--30\%. For
example, an error of this magnitude is claimed to be needed to make
the directly imaged extrasolar planets around HR~8799 dynamically
stable as their model-derived masses seem to be $\sim$30\% too high
\cite{fm-8799}.  However, we emphasize that this apparent
over-luminosity is based on a single system whose age is estimated
from the primary star HD~130948A.  While this young solar analog is
fortuitously amenable to multiple age-dating techniques using stellar
rotation, chromospheric and x-ray activity, and isochrone fitting,
precise age estimation is notoriously difficult for an arbitrary field
star.  Thus, refining the age estimate for HD~130948A (e.g., via
asteroseismology) is essential.  Also, more dynamical masses for the
substellar binaries in triple systems with stars of known age (e.g.,
$\epsilon$~Ind~Bab, Gl~417BC, GJ~1001BC) are critically needed to
address this potential problem with model cooling rates.

\section{Future Work}
\label{sec:future}

To make progress in understanding the problems we have found with
substellar models, we are in the process of developing a larger sample
of masses spanning a wide range in temperature, mass, and age.  For
example, more mass measurements should determine whether the observed
over-luminosity of brown dwarfs (Figure~\ref{fig:lbol-age}) persists
for different surface temperatures.  If so, this would point to a
problem with the interior structure model (e.g., convection) rather
than a surface effect (e.g., magnetic fields).

Resolved spectroscopy is also needed to enable detailed
characterization of the individual components' temperatures, surface
gravities, and abundances.  Binaries with known masses can provide
stronger tests of atmospheric models than field brown dwarfs of
unknown age or mass.  Such measurements will be able to assess poorly
understood atmospheric effects such as dust formation and
sedimentation (e.g., \cite{2001ApJ...556..357A}) and vertical mixing
that can drive nonequilibrium chemistry (e.g.,
\cite{2006ApJ...647..552S}).

\begin{figure}

\centerline{
%\resizebox{0.9\columnwidth}{!}{\includegraphics{plot_lithium_age-bw-cs15.ps}}
\resizebox{0.9\columnwidth}{!}{\includegraphics{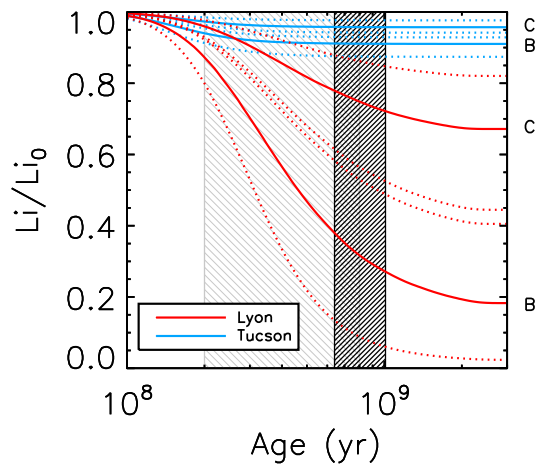}}
}
\caption{Lithium depletion as a function of age predicted by two
  independent sets of evolutionary models.  The solid lines correspond
  to the individual masses of HD~130948B and C.  These lines are
  bracketed by dotted lines that correspond to the 1$\sigma$
  uncertainties in the individual masses.  The ordinate is the
  fraction of initial lithium remaining.  The hatched black box
  indicates the constraint from the age of the primary star HD~130948A
  estimated from gyrochronology (gray is the less precise but
  consistent estimate from chromospheric activity).  The Tucson models
  predict that both of the binary components should be
  lithium-bearing, while the Lyon models predict that at least one of
  the two components should be massive enough to have destroyed its
  initial lithium.  Thus, resolved optical spectroscopy of lithium
  absorption at 6708~\AA\ for this binary, only possible with
  \HST/STIS, will provide the first test of the model-predicted
  lithium depletion boundary, which has been widely used, for example,
  to derive the canonical ages for young clusters such as the Pleiades
  \cite{stauf98}.}
\label{fig:lith}

\end{figure}

In addition, optical spectra from \HST/STIS will enable measurements
of lithium absorption at 6708~\AA\ for binaries that we have shown are
very close to the theoretically predicted lithium burning limit at
$\approx$60~\Mjup, providing a novel test of substellar interior
models (\cite{2009ApJ...692..729D}, \cite{me-2397a}).  As shown in
Figure~\ref{fig:lith}, independent groups make different predictions
for the mass limit of lithium depletion.  Binaries close to this limit
offer the chance to empirically determine the lithium boundary for the
first time, with a precision comparable to our dynamical mass
measurements ($\sim$3\%).  The theoretically predicted lithium
boundary has never been directly tested in such a manner, even though
it has been widely used, for example, to determine the canonical ages
for young clusters such as the Pleiades \cite{stauf98}.

Finally, a large sample of ultracool binary orbits will provide
constraints on formation models for low-mass stars and brown dwarfs as
the orbital parameters of binaries (e.g., semimajor axis and
eccentricity) record the dynamical imprint of their formation process.
For example, competing models predict very different eccentricity
distributions for ultracool binaries: Stamatellos \& Whitworth predict
high eccentricities ($e \gtrsim 0.5$) for ultracool binaries formed
via gravitational fragmentation in a massive circumstellar disk
\cite{2009MNRAS.392..413S}, while Bate predicts modest eccentricities
($e \lesssim 0.5$) for ultracool binaries formed in a cluster
environment \cite{2009MNRAS.392..590B}.

\end{document}